\documentstyle[11pt,newpasp,twoside,epsf]{article}
\def\edcomment#1{\iffalse\marginpar{\raggedright\sl#1\/}\else\relax\fi}
\reversemarginpar

\begin{document}
\title{On the Stability of Particles in Extrasolar Planetary Systems}
  \author{Eugenio J.\ Rivera \& Nader Haghighipour}
\affil{Carnegie Institution of Washington, Dept.\ of Terrestrial Magnetism, 
5241 Broad Branch Rd NW, Washington, DC 20015}

\begin{abstract}
In this paper, we present preliminary results on the stability of massless
particles in two and three-planet systems.  The results of our study may be
used to address questions concerning the stability of terrestrial planets in
these systems and also the trapping of particles in resonances with the
planets.  The possibility of the existence of islands of stability and/or
instability at different regions in multi-body systems and their probable
correspondence to certain resonances are also discussed.
\end{abstract}

\section{Introduction}
The discovery of multi-body extrasolar planetary systems during the past few
years has once again confronted astrodynamicists with the old question of the
stability of such systems.  The discovery of GJ 876 where two planets are
locked in a 2:1 mean-motion resonance (Marcy et al.\ 2001), the
confirmation of three planets in orbit about Upsilon Andromedae (Butler et al.\
1999), and the discovery of planetary systems around 47 Uma and 55 Cancri with
planets in orbits more closely resembling those in the Solar System
(Fischer et al.\ 2002; Marcy et al.\ 2002), have set the grounds for a deeper
look at the problem of the stability of multi-body systems.  Here we present a
preliminary investigation of the dynamical stability of the 47 Uma and
55 Cancri systems.  A more detailed analysis, including similar work on GJ 876
and $\upsilon$ And, will be addressed in future work.

\section{Methodology}
We performed dynamical fits to the radial velocity data for the stars
47 Uma and 55 Cancri using a Levenberg-Marquardt minimization
algorithm (Press et al.\ 1992), as in Laughlin \& Chambers (2001) and
Rivera \& Lissauer (2001).  We assumed that the planets in each system
were coplanar and that the plane of the planets contained the line of
sight.  We then used the resulting fitted parameters as initial
conditions for N-body simulations.  In order to give the resulting systems
the ability to explore extra degrees of freedom, we artificially added a
mutual inclination of one degree.  
We also performed simulations of each system in which hundreds of (massless)
test particles were added.  All test particles were started on circular orbits
with respect to the central star.  No test particles were placed in regions
such that a test particle would initially cross the orbit of a planet.  The
results may be used as an indication of the presence and locations of potential
terrestrial planets in these systems.

The simulations were performed with the second-order mixed variable
symplectic (MVS) integrator in the MERCURY integration package
(Chambers 1999), which is based on the technique pioneered by
Wisdom \& Holman (1991).  This code was modified to include the
principal effects of general relativity, as in Lissauer \& Rivera (2001).

\section{47 Uma}
In agreement with Fischer et al.\ (2002), we find that in dynamical fits
to the radial velocity data for 47 Uma, the value of $\chi^2_{\nu}$ does
not change significantly for a fit with the eccentricity of the outer
companion in the range 0 to 0.2.  When we fit for the eccentricity
of the outer planet, the fitting routine converged on a value of 0.19.
Note that this value is near the boundary of stability determined by
Fischer et al.\ (2002).  A long-term simulation of the system based on this
fit is stable over billions of years.  In this simulation, the two
companions are in a 5:2 mean motion resonance.
Also, throughout the simulation the orbits are nearly anti-aligned.  Table 1
gives the parameters from the fit at epoch JD 2446959.737, and Figure 1a
shows the nearest and farthest points in the companions' orbits vs.\ time
for the first 1 Gyr of a long-term simulation based on the parameters
given in Table 1.  Figure 2 shows the eccentricities, inclinations, and the
period ratio of the companions over short times.

Figure 1b shows the stability of test particles in a 10 Myr simulation of
the same system.  It shows the time that a test particle was lost
vs.\ its initial semimajor axis.  The dots and lines toward the
bottom of the figure indicate the initial semimajor axes and radial
excursions of the planets.  Note that islands of instability occur
at the locations of several mean motion resonances with the planets.
Also, exterior to the outer companion, there are islands of
stability between mean motion resonances.  Since 47 Uma is similar to
the sun, the region around 1 AU is of particular interest.  Unfortunately,
the 3:1 mean motion resonance with the inner companion lies in this region,
and there are signs that the region may not be stable.  Note that while
only the one test particle at 1 AU was lost in less than 10 Myr, proximity
to the 3:1 mean motion resonance could endanger the stability of other
nearby orbits over longer timescales.

\begin{table}[ht]
\caption{Astrocentric orbital Parameters for the 47 Uma and 55 Cancri systems}
\begin{center}
\begin{tabular}{lrrcrrr}
\tableline
 & \multicolumn{2}{c}{47 Uma} & \phantom{x} & \multicolumn{3}{c}{55 Cancri}\\
Parameter & inner & outer & \phantom{x} & inner & middle & outer\\
\tableline
Period (days)                & 1082 & 2735 & \phantom{x} & 14.65 & 44.58 & 5582\\
$e$                          & 0.034 & 0.190 & \phantom{x} & 0.021 & 0.169 & 0.141\\
$\omega$ (deg)               & 133 & 284 & \phantom{x} & 104 &  56 & 202\\
Mean Anomaly (deg)           & 351 & 329 & \phantom{x} & 340 & 345 & 189\\
$m_{\rm pl}$ ($M_{\rm Jup}$) & 2.65 & 0.90 & \phantom{x} & 0.84 & 0.20 & 4.21\\
\tableline
\tableline
\end{tabular}
\end{center}
\end{table}

\begin{figure}[ht]
\plotone{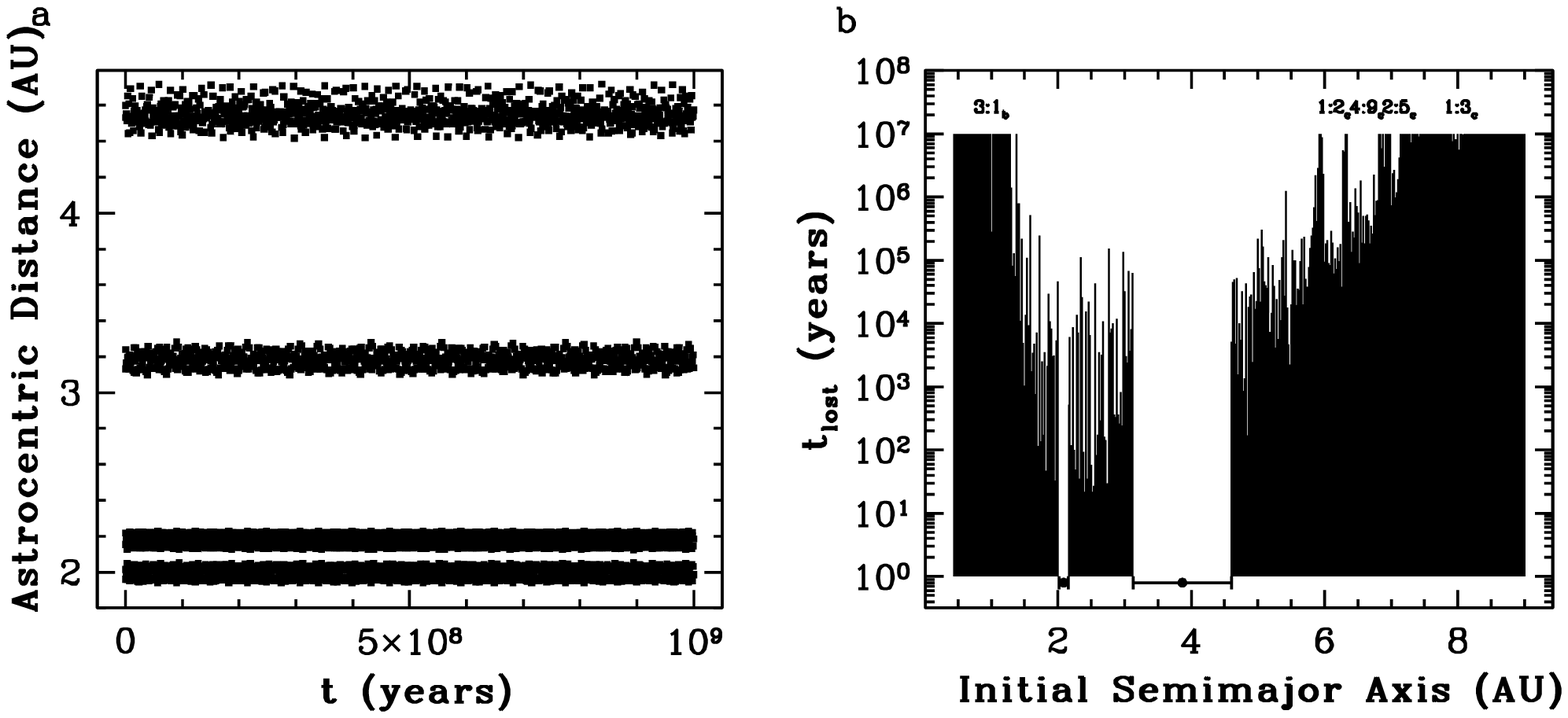}
\caption{(a) Periastra and apastra of the companions of 47 Uma.
(b) Stability times of test particles in the 47 Uma system.}
\vspace{22pt}
\plotone{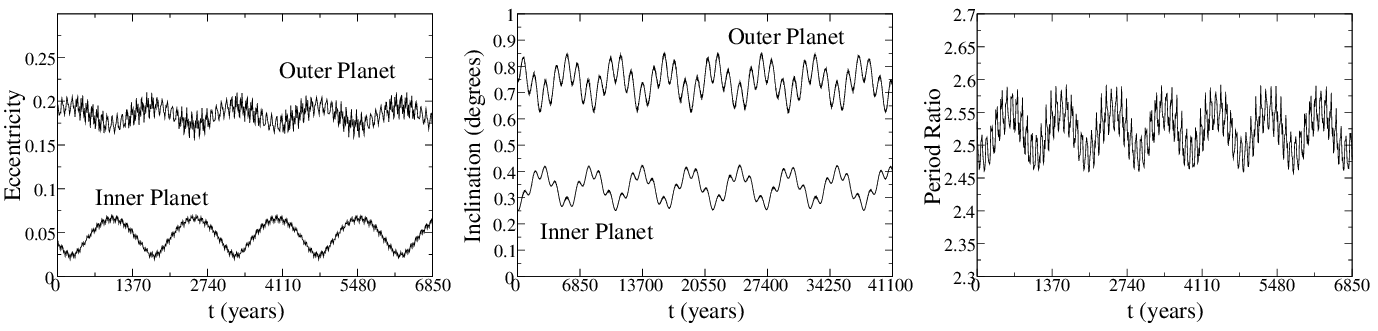}
\caption{Eccentricities, inclinations, and period ratio of the companions of
47 Uma.}
\end{figure}

\section{55 Cancri}
This system proved to be difficult to fit with the Levenberg-Marquardt
routine.  We obtained two fits with almost the same $\chi^2_{\nu}$ value
with very different parameters for the outer planet.  This is an
indication of the large uncertainties in this planet's orbital
parameters.  Table 1 gives the parameters from one of our fits at
epoch JD 2450250.0, and Figure 3a shows the periastron and apastron of each
planet vs.\ time from a simulation based on this fit.  The simulation
shows that the inner and middle companions are near a 3:1 mean motion
resonance.  These results are in agreement with Marcy at al.\ (2002).

Figure 3b shows the stability of test particles in a 5 Myr simulation of
the same system.  It clearly shows a large stable region between the
middle and outer companions.  This is in rough agreement with
Marcy et al.\ (2002), in which they showed that a terrestrial planet
on a circular orbit at 1 AU would be stable.

\begin{figure}[ht]
\plotone{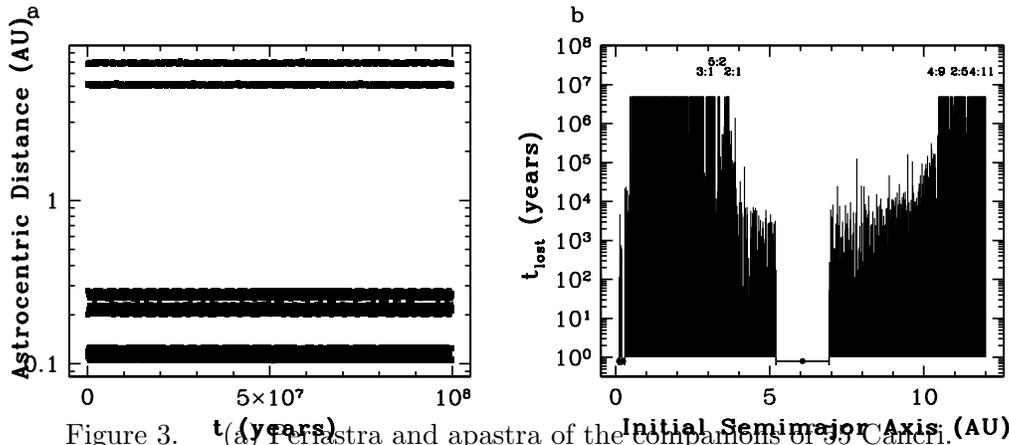}
\vspace{-22pt}
\caption{(a) Periastra and apastra of the companions of 55 Cancri.
(b) Stability times of test particles in the 55 Cancri system.}
\end{figure}

\section{Summary}
We have presented results on the stability of the 47 Uma and 55 Cancri systems
with and without massless particles.  Dynamical fits were used to determine
the initial conditions for the simulations.  A large
area of parameter space is consistent with the observations.  The area
around 1 AU in the 47 Uma system may not harbor a terrestrial planet, while
the opposite is true in the 55 Cancri system.  These results are in rough
agreement with previous studies (Fischer et al.\ 2002; Marcy et al.\ 2002).

\acknowledgments
We thank Debra Fischer, Geoff Marcy, and Paul Butler for providing their radial
velocity observations.

\end{document}